\documentclass[twocolumn,showpacs,preprintnumbers,amsmath,amssymb]{revtex4}
\usepackage{graphicx, epsfig, psfrag}

\def\qq{$ q\bar q $}  
\def\ss{$ s\bar s $}  
\def\nn{$n\bar{n}$}  
\def\uu{$u\bar u$}  
\def\dd{$d\bar d$}

\begin{document}   
   
\title{The ``forbidden" decays of hybrid mesons to $\pi \rho$ can be large} 
\author{F.E.Close}  
\email{F.Close1@physics.ox.ac.uk}  
\author{J.J.Dudek}  
\email{dudek@thphys.ox.ac.uk} 
\affiliation{Department of Physics - Theoretical Physics, University of Oxford,\\ 
1 Keble Rd., Oxford OX1 3NP, UK} 
   
\begin{abstract}   
  The observation of $\pi_1(1600) \to \pi \rho$ is shown in the
  flux-tube model to be compatible with this state being a hybrid
  meson with branching ratio to this channel $\sim 30\%$.  The $\pi \rho$ widths of other hybrids are related by rather
  general arguments. These results enable cross sections for
  photoproduction of hybrids to be predicted.
\end{abstract}   
\maketitle

The ``smoking gun" for hybrid mesons has been the possibility of them
having combinations of $J^{PC}$ that are forbidden to conventional
$q\bar{q}$.  Examples of such exotic states include $1^{-+}$, which in
lattice QCD and the flux-tube model is predicted to occur $\sim 2$ GeV
in mass\cite{hybrids,IP,bcs}, with the further exotic states
$0^{+-},2^{+-}$ occurring somewhat higher in mass\cite{hybrids}.
 
A $1^{-+}$ state, $\pi_1(1600)$, has been claimed in independent
experiments\cite{e852,Lu:2004yn, oneminusplus}.  Its mass is somewhat lower
than lattice computations had anticipated, but allowing for the
lighter masses of \nn ($\equiv$ \uu~ or \dd~) relative to the \ss~
states studied in ref.\cite{hybrids}, a $1^{-+}$ with a mass $\sim
1780(200)$MeV is not implausible \cite{michael2000}. These observations are in different channels and it has not yet been established that they refer to a single state\cite{indiana}. Observation in the channel $\pi \rho$\cite{e852} has raised
questions about its nature, given that the standard predictions of
hadronic decays of hybrids have been that they decay into excited
states\cite{IK,CP95} and are even forbidden into $\pi
\rho$\cite{IK,CP95,page}.
 
The latter selection rule applies in a symmetry limit (specifically
where the $\pi$ and $\rho$ have the same size) and in the case where
the decay is triggered by breaking the flux tube. It is not a general
axiom. The approach described in the present paper severely breaks
this symmetry by shrinking the $\pi$ to a point-like current.  This
follows a standard approach for calculating pion emission, which has
been applied with reasonable success to conventional decays for over
30 years\cite{fkr,Godfrey:1985xj,3p0model}. We are now able to apply it to the
decays of hybrids following the recent development in
ref.\cite{CD03a,CD03b}. This built on an insight of
Isgur\cite{isgur99}, which in simplistic terms is that the flux-tube
is a dynamical degree of freedom which can be excited by the action of
a current on its ends - the quarks. (The application of this idea is
described extensively in our paper \cite{CD03b}). It is the purpose of
the present paper to apply these ideas to calculate $\pi \rho$ decays
of hybrids. We shall see that they can be large. Our results reveal
that existing calculations in the literature implicitly allow this,
and that the $J^{PC}$ dependence of our results is also found to occur
in those calculations.  Finally, given the empirical success of
converting $\pi \rho$ amplitudes to $\pi \gamma$ for known states, we
can predict the $\gamma \pi \to \cal{H}$ amplitudes, which are an
essential requirement for estimating their photoproduction cross
sections.

\section*{Pion emission from conventional mesons} 
Emission of a $\pi^+$ by the quark in a meson ($q_i \bar{q}$) has matrix element
{ \small
\begin{multline}
{\cal M}(q_i \to q_f + \pi^+) = \\
\int d^3 \vec{p} \int d^3 \vec{p}' \phi^*_f(\vec{p}')
\phi_i(\vec{p}) \; \left \langle \bar{q} (\tfrac{1}{2} \vec{P}_f + \vec{p}') \right|\left.
\bar{q}(\tfrac{1}{2} \vec{P}_i + \vec{p}) \right \rangle \\
\times \left \langle q_f (\tfrac{1}{2} \vec{P}_f - \vec{p}') \right| \tfrac{g}{2m} \int d^3 \vec{r} \; \bar{\psi}(\vec{r})  \gamma^\mu q_\mu \gamma^5 \tfrac{\tau^-}{\sqrt{2}} \psi(\vec{r}) e^{- i \vec{q}.\vec{r}} \left| q_i (\tfrac{1}{2} \vec{P}_i - \vec{p}) \right\rangle, \nonumber
\end{multline}
}
where the structure of the current is the divergence of the axial current, but to the order in the non-relativistic reduction that we will work could equally well be a pseudoscalar form;
$\tfrac{\tau^-}{\sqrt{2}}$ is the isospin lowering operator, and the
$\phi(\vec{p})$ are the internal momentum wavefunctions. $g$ is the
pion-quark-quark coupling constant which we will determine by fitting
conventional meson decay rates.

Expanding the Dirac spinors $\psi$ in terms of quark creation and
annihilation operators, retaining only those which will annihilate
$q_i$ and create $q_f$ and performing the non-relativistic reduction working in the rest frame of the initial meson we find
\begin{multline}
  {\cal M}(q_i \to q_f + \pi^+) =  - \tfrac{g}{2m} F(q_i, q_f) \int
  d^3 \vec{p} \; \phi^*_f(\vec{p} + \vec{q}/2) \phi_i(\vec{p}) \\
\times \big[ \vec{\sigma}\cdot \vec{q} (1+ \tfrac{q^0}{2 m}) + \tfrac{q^0}{m} \vec{\sigma} \cdot \vec{p} + \ldots \big].\nonumber
\end{multline}
where $F(q_i, q_f)= \langle q_f| \tfrac{\tau^-}{\sqrt{2}}| q_i \rangle$ is a flavour factor accounting for isospin
conservation at the pion-quark-quark vertex. Fourier
transforming with $\phi(\vec{p}) = \int d^3 \vec{r} e^{- i
  \vec{p}.\vec{r}} \psi(\vec{r})$ gives
\begin{multline} \label{convpion}
  {\cal M}(\substack{q_i \\\bar{q}_i} \to \substack{q_f \\ \bar{q}_f}
  \pi^+) =  \mp \tfrac{g}{2m} F(\substack{q_i \\\bar{q}_i},
  \substack{q_f \\ \bar{q}_f}) \\
\times \int d^3 \vec{r} \; \psi^*_f(\vec{r})
  \big[ \vec{\sigma}\cdot \vec{q}  \mp \tfrac{q^0}{m} \vec{\sigma} \cdot \vec{p}'\big] e^{\pm i \vec{q}.\vec{r}/2} \psi_i(\vec{r}) 
\end{multline}
for emission by the quark or the antiquark and where the operator $\vec{p}'$ is $-i \vec{\nabla}_{\vec{r}}$ acting
backwards onto the final state wavefunction.

Compare equation \eqref{convpion} with equation (19) in \cite{Godfrey:1985xj},
which has the opposite sign definition for $\vec{r}$. Their decay
widths are in terms of two independent form-factor parameters, $g,h$
which they fit to data. Our approach determines for example $h$ in
terms of $g$ and we hence have less freedom to fit, but more
predictive power.
 
Using the representation {\footnotesize $u \sim \begin{pmatrix} 1 \\0 \end{pmatrix}$,  $d \sim \begin{pmatrix}
  0 \\1 \end{pmatrix}$, $\bar{u} \sim \begin{pmatrix} 0 \\1 \end{pmatrix}$, $\bar{d} \sim \begin{pmatrix}
  -1 \\0 \end{pmatrix} $} and the explicit form {\footnotesize
$\tfrac{\tau^-}{\sqrt{2}} \sim \begin{pmatrix} 0 & 0 \\1&0
\end{pmatrix}$} we see that the only non-zero flavour factors are
$F(d,u) = +1$ and $F(\bar{u}, \bar{d})=-1$. We find that equation
\eqref{convpion} with these charge factors correctly conserves isospin
and $G$-parity.

With our normalisations, the partial width of a meson of spin-$J$ is
given by
\begin{equation}\label{widthpirho}
  \Gamma(M_J \to \pi V) = \frac{C}{2J+1} \frac{|\vec{q}|}{2 \pi} \sum |{\cal M}|^2,
\end{equation}
where the sum is over quark and anti-quark emission and either
helicities or partial waves. $C$ is the number of end-state charge
possibilities (e.g. $C=2$ for isovector to $\rho \pi$, $C=1$ for
isovector to $\omega \pi$).

We now have all the tools required to calculate decay widths. We will
demonstrate the method with the important channel $b_1 \to \omega
\pi$.

\section*{ $b_1 \to \omega \pi$ in the pion emission model}

Considering first the spin and spatial dependence and leaving for now
the flavour dependence we can write for the amplitude, $A_{q,\bar{q}}=$
\begin{multline}\label{}
\hspace{-3mm}\int\!\!\! r^2 dr \!\! \int\!\!\! d \Omega \;
  \frac{R_f^*(r)}{\sqrt{4 \pi}}  \left\langle S=1, m_s \left| \big[
  \vec{\sigma}_{q, \bar{q}} \cdot \vec{q} \mp \tfrac{q^0}{m}
  \vec{\sigma}_{q, \bar{q}} \cdot \vec{p}'\big] \right|S=0 \right\rangle \\
\times e^{\pm i \vec{q}\cdot \vec{r}/2} Y^{m_L}_1(\Omega) R_i(r). \nonumber
\end{multline}
Evaluating the angular integrals and the spin matrix element gives us the helicity amplitudes,
\begin{align}
    A_{q,\bar{q}}(0) &= i \frac{g}{2m} \frac{|\vec{q}| }{\sqrt{3}}  \left[ 3
      \langle j_1 \rangle + \frac{\langle \overleftarrow{\partial} j_0
        \rangle}{m} - 2 \frac{\langle \overleftarrow{\partial} j_2
        \rangle}{m} \right] \nonumber\\
    A_{q,\bar{q}}(\pm) &= i \frac{g}{2m} \frac{|\vec{q}| }{\sqrt{3}} \left[ \frac{\langle \overleftarrow{\partial} j_0
        \rangle}{m} + \frac{\langle \overleftarrow{\partial} j_2
        \rangle}{m} \right],\nonumber
\end{align} 
where $\langle \overleftarrow{\partial} j_L \rangle$ is shorthand for
$\int r^2 dr \tfrac{d R_f^*}{dr} j_L(|\vec{q}|r/2) R_i$.

Partial wave amplitudes can be constructed according to Table XI of
\cite{Godfrey:1985xj}. We find 
\begin{align}
    A_{q,\bar{q}}(S) &= i \frac{g}{2m} |\vec{q}|   \left[
      \langle j_1 \rangle + \frac{\langle \overleftarrow{\partial} j_0
        \rangle}{m} \right]\nonumber \\
    A_{q,\bar{q}}(D) &= -i \sqrt{2} \frac{g}{2m} |\vec{q}|   \left[  \langle j_1 \rangle  - \frac{\langle \overleftarrow{\partial} j_2
        \rangle}{m} \right].\nonumber
\end{align} 
Including the flavour factors for the decay of $b_1^+$ we obtain a
non-zero matrix element for the end state $\omega \pi^+$ only, and not for example, $\rho^+ \pi^0$, in line
with conservation of isospin and $G$-parity. The $D/S$ amplitude ratio
then for this decay is
\begin{equation}\label{dtos}
  \frac{D}{S}(b_1 \to \omega \pi) = - \sqrt{2} \frac{\langle j_1 \rangle - \frac{\langle \overleftarrow{\partial} j_2
        \rangle}{m}}{ \langle j_1 \rangle + \frac{\langle \overleftarrow{\partial} j_0
        \rangle}{m} },
\end{equation}

A sensitive test of hadron decay models comes from comparing the $D/S$
amplitude ratios for the decays $b_1 \to \omega \pi, a_1 \to \rho \pi$
with the experiment world averages $+0.277(27), -0.108(16)$\cite{Hagiwara:2002fs}. For
the $a_1 \to \rho \pi$ decay in this model we obtain $\frac{D}{S}(a_1 \to \rho \pi) = - \frac{1}{2} \frac{D}{S}(b_1 \to \omega \pi)$, which is also found in the $^3 P_0$ model and the
``$sKs$'' and ``$j^0 K j^0$'' models discussed in \cite{Ackleh:1996yt}.

Using wavefunctions obtained variationally from the Isgur-Paton meson
Hamiltonian (``IP'') (see Appendix D of \cite{CD03b}) we obtain ratios
$+0.45$ and $-0.22$ for $\frac{D}{S}(b_1 \to \omega \pi),
\frac{D}{S}(a_1 \to \rho \pi)$, which have the right sign and relative size, but are
roughly a factor of 2 too large in magnitude.  A standard
approximation \cite{Ackleh:1996yt} is to describe the mesons by
harmonic oscillator wavefunctions with a single $\beta$ value for all
states. If we do this and fit to the experimental $b_1$ ratio we find
$\beta = 0.39(3)$GeV, which is considerably larger than $\bar{\beta}
\approx 0.31$GeV in ``IP'' but is in good agreement with $\beta
\approx 0.4$ in \cite{Ackleh:1996yt}.

That the effective $\beta$ is larger than our ``IP'' value may be due
to our pointlike pion approximation. Experimentally the pion is not
pointlike, it has a charge radius comparable with other light mesons
and our radial wavefunction overlap should really take some account of
this. Quite possibly we are feeling this in the increased $\beta$,
which has subsumed the effect of the pion wavefunction. However it is
worth noting that the effective $\beta$ values (presented in
\cite{IK}) for the meson wavefunctions computed in the model of
Godfrey and Isgur\cite{Godfrey:1985xj} are much larger than for the ``IP''
Hamiltonian, and much closer to the $\beta=0.4$ GeV preferred above. Godfrey and Isgur use
a partially relativised Hamiltonian and consider spin-dependent terms,
so it may be that the simple ``IP'' Hamiltonian as applied to
conventional mesons is missing some important effects which set the
size of meson states.

\section*{Conventional meson decays to $\pi V$}\label{sect:convdecay}

\begin{table}
\begin{center}
\begin{tabular}{l|l}
$S$-waves & $D$-waves\\
$\Gamma_S(a_1 \to \rho \pi) = \frac{8}{3} \Sigma$ & $\Gamma_D(a_1 \to \rho \pi) = \frac{4}{3} \Delta$ \\
$\Gamma_S(b_1 \to \omega \pi) = \frac{2}{3} \Sigma$ & $\Gamma_D(b_1 \to \omega \pi) = \frac{4}{3} \Delta$ \\
-& $\Gamma_D(a_2 \to \rho \pi) = \frac{12}{5} \Delta$ \\
\hline
$P$-waves & $F$-waves \\
$\Gamma_P(\pi_2 \to \rho \pi) = \frac{8}{5} \Pi $ & $\Gamma_F(\pi_2 \to \rho \pi) = \frac{12}{5} \Phi$ \\
- & $\Gamma_F(\omega_3 \to \rho \pi) = \frac{24}{7} \Phi$\\
- & $\Gamma_F(\rho_3 \to \omega \pi) = \frac{8}{7} \Phi$\\
\end{tabular}
\begin{tabular}{l}
$\Sigma \equiv \frac{|\vec{q}|^3}{2 \pi}\left(\frac{g}{2m}\right)^2 
\left(\langle j_1 \rangle + \frac{\langle \overleftarrow{\partial} j_0 \rangle}{m}\right)^2 $ \\ 
$\Delta \equiv \frac{|\vec{q}|^3}{2 \pi}\left(\frac{g}{2m}\right)^2 
\left(\langle j_1 \rangle - \frac{\langle \overleftarrow{\partial} j_2 \rangle}{m}\right)^2 $ \\
$\Pi \equiv \frac{|\vec{q}|^3}{2 \pi}\left(\frac{g}{2m}\right)^2 
\left(\langle j_2 \rangle + \frac{\langle \overleftarrow{\partial} j_1 \rangle}{m}\right)^2 $ \\ 
$\Phi \equiv \frac{|\vec{q}|^3}{2 \pi}\left(\frac{g}{2m}\right)^2
\left(\langle j_2 \rangle - \frac{\langle \overleftarrow{\partial} j_3 \rangle}{m}\right)^2 $ \\
\end{tabular}
\end{center}
\caption[Conventional meson $\pi V$ decay widths]{$\pi V$ decay widths for $L=1,2$ conventional light-quark mesons in the pion emission model. \label{tab:convpion}}
\end{table}

The $\pi \substack{\rho \\ \omega}$ decays of the spin-triplet $L=1,2$
mesons can be computed in this model, the results being tabulated in
Table \ref{tab:convpion}. Another precision test of the decay model is the
$F/P$ ratio of the $\pi_2(1670) \to \rho \pi$ decay, which has recently been measured for the first time by the E852 collaboration who find $F/P = -0.72 \pm 0.07 \pm 0.14$\cite{Chung:1999we}. In the pion emission model we find
\begin{equation}\label{}
  \frac{F}{P}= - \sqrt{\frac{3}{2}} \frac{\langle j_2 \rangle -
    \frac{\langle \overleftarrow{\partial} j_3 \rangle}{m}}{\langle
    j_2 \rangle +\frac{\langle \overleftarrow{\partial} j_1
      \rangle}{m} }.\nonumber
\end{equation}
With ``IP'', $\beta=0.4$ wavefunctions this would equal $+0.57, +0.31$,
neither of which is compatible with the experimental value. Since
this is an $L=2 \to L=0$ transition we might expect there to be a
different effective $\beta$, which we can fit using the experimental value. With equal $\beta$ harmonic oscillator
wavefunctions we find
\begin{equation}\label{}
  \frac{F}{P} = - \sqrt{\frac{3}{2}}\frac{1+ \frac{|\vec{q}|}{4m}}{1+ \frac{|\vec{q}|}{4m} - \frac{10 \beta^2}{|\vec{q}| m}},
\end{equation}
which cannot be satisfied by any real $\beta$ with the experimental
numbers. So we see that the pion emission model as formulated here
cannot accommodate the E852 value for $F/P$. The $^3 P_0$ decay model successfully accommodates the $D/S$ ratios discussed earlier with
parameter values which do a good job of describing a wide range of
hadron decays. With the same parameters it predicts $F/P \sim +0.4$\cite{Barnescomm}
which is in the same region as the pion emission predictions and not
compatible with experiment. If the E852 result is confirmed it casts doubt over the validity
of these commonly used non-relativistic hadron decay models for high partial waves. 

We present in Table \ref{pionwidths} the partial widths for a number
of meson decays computed in the model using $g=2,3$ for $\beta=0.4$
wavefunctions and $g=3$ for ``IP'' wavefunctions. Also shown are the
predictions of the $^3 P_0$ model taken from \cite{Ackleh:1996yt, Barnes:1997ff} and
experimental values taken from \cite{Hagiwara:2002fs}.

\begin{table}
\begin{tabular}{c|cccc} 
Mode & ``IP'' $(g=3)$ & $\beta=0.4 (g=2,3)$ & $^3P_0$ &  Data \\ 
\hline 
$\Gamma_S(a_1 \to \rho \pi)$ &  280 & (255, 580) & 530 & 150 $\to$ 360\\
$\Gamma_D(a_1 \to \rho \pi)$ & 14 & (5,10) & 15 & 3 $\to$ 8\\
$\Gamma_S(b_1 \to \omega \pi)$ & 70 & (64,145) & 132 & $<$ 132\\ 
$\Gamma_D(b_1 \to \omega \pi)$ & 14 & (5,10) & 11 & $<$ 10 \\ 
$\Gamma_D(a_2 \to \rho \pi)$ & 52 & (18,40) & 54 &75 $\pm$ 7\\ 
$\Gamma_{F+P}(\pi_2 \to \rho \pi)$ & 162 &(131, 297)& 118 & 81 $\pm$ 11\\
$\Gamma_F(\omega_3 \to \rho \pi)$ & 77 &(16,36)&50 & $<$ 74\\
$\Gamma_F(\rho_3 \to \omega \pi)$ & 19& (5,12)& 19 & 26 $\pm$ 13 
\end{tabular} 
\caption[Numerical estimates of conventional meson $\pi V$ decay widths]{Numerical estimates of  $\pi V$ decay widths in MeV for $L=1,2$ conventional light-quark mesons in the pion emission model using wavefunction parameterisations as described in the text. \label{pionwidths}}
\end{table}

Unfortunately there is little precision data on $\pi \substack{\rho \\
  \omega}$ decays available, so the best we can say is that the pion
emission model does a reasonable job of describing the data. The
qualitative pattern of large, small and intermediate empirical widths
is faithfully reproduced which suggests that the underlying group
transformation properties of the states and the pion transition
operator are valid. We cannot
accurately pin down the value of the coupling using the experimental
data. Note also that despite their failure to predict the precision
$D/S$ ratios, the ``IP'' wavefunctions do as good a job overall of describing
the data as the $\beta=0.4$ wavefunctions.

\section*{Hybrid meson decays to $\pi V$}

We can derive the matrix element for decay in the same manner as we
did for conventional decays, but now explicitly including the
flux-tube degrees-of-freedom. The essential change lies in the
modification of the quark positions and momenta as detailed in
\cite{CD03a,CD03b},
\begin{eqnarray}
 \vec{r}_{q, \bar{q}} =
\pm \frac{1}{2}\vec{r} - \sqrt{\frac{2b}{\pi^3}} \frac{\beta_1 r}{m}
\vec{a}; \\
\vec{p}_{q, \bar{q}} = \pm \vec{p} - i\sqrt{\frac{2b}{\pi^3}} \pi\beta_1 \vec{a};
\label{define} 
\end{eqnarray} 
where $\vec{r}$ is the internal ``longitudinal" relative coordinate
through the c.m. and parallel to the \qq~ axis, and $\vec{a}$ is the
transverse Fourier mode of the flux tube associated with the
transverse displacement of the \qq~ relative to the
$\vec{r}$\cite{isgur99,CD03b}. We have already used the fact that $\vec{r}$ can
cause transitions between $S$ and $P$ wave conventional \qq~ states;
$\vec{a}$ can analogously generate transitions between
conventional and hybrid (excited flux-tube) states. The parameter
$\beta_1$ is effectively the measure of the transverse extent of the
flux-tube wavefunction which will not appear in state widths; if
described by Gaussian wavefunctions\cite{isgur99,CD03b} one
effectively has
\begin{equation}
 \vec{p}_a |g.s.\rangle = i\beta_1^2 \vec{a}| g.s. \rangle
\nonumber 
\end{equation}
which has been used in deriving the expression in eq.(\ref{define}).

The matrix element for the lightest hybrid multiplet (one phonon in the $p=1$ mode) to decay to $\pi\rho$, retaining only the required terms linear in $\vec{a}$ is,
\begin{multline}\label{hybridcurrent}
 {\cal M}_{q, \bar{q}} =  \mp  i \tfrac{g}{2m} \sqrt{\tfrac{2 b}{\pi}} \beta_1 F(\substack{q_i
   \\\bar{q}_i}, \substack{q_f \\ \bar{q}_f})
 \int d^3 \vec{r} \int d^2 \vec{a}
 \psi^{({\cal C})*}_{\rho}(\vec{r}) \chi^*_0(\vec{a}) \\
\times \left\langle S=1, m_\rho \left| \left(\vec{\sigma}_{q,
        \bar{q}}. \vec{q} \mp \frac{q^0}{m} \vec{\sigma}_{q,
        \bar{q}}.\vec{p}'  \right) \frac{r}{\pi m}\vec{q}.\vec{a} -
    \frac{q^0}{m} \vec{\sigma}_{q, \bar{q}}.\vec{a} \right|S, m_s'
\right\rangle \\
\times e^{\pm i \vec{q}.\vec{r}/2} \psi^{({\cal H})}_{m'}(\vec{r}) \chi_1(\vec{a}).
\end{multline}
The $\vec{\sigma}.\vec{p}' \vec{q}.\vec{a}$ term is formally
suppressed at order $|\vec{q}|/m$ relative to the leading term and as
such we will neglect it initially. We will return at the end of this paper to consider the effect it and other neglected terms might
have.

The calculation of matrix elements can be performed immediately using
the techniques of ref.\cite{CD03b} (see especially eqns.(18,19) of
that paper). We present the results in Table \ref{pionhyb}.

\begin{table}[h]
\hspace{-1cm}
\begin{tabular}{l|l}
\multicolumn{2}{c}{${\cal P_H}=+$} \\
$S$-waves & $D$-waves\\
$\Gamma_S(a_{1H} \to \rho \pi) = \frac{1}{3} \Sigma_{\cal H}$ & $\Gamma_D(a_{1H} \to \rho \pi) = \frac{1}{6} \Delta_{\cal H}$ \\
$\Gamma_S(b_{1H} \to \omega \pi) = \frac{1}{3} \Sigma_{\cal H}$ & $\Gamma_D(b_{1H} \to \omega \pi) = \frac{1}{24} \Delta_{\cal H}$ \\
-& $\Gamma_D(b_{2H} \to \omega \pi) = \frac{3}{40} \Delta_{\cal H}$ \\
\hline
\multicolumn{2}{c}{${\cal P_H}=-$}\\
 $P$-waves & $F$-waves \\
$\Gamma_P(\rho_H \to \omega \pi) = \frac{1}{4} \Pi_{\cal H} $ & - \\
$\Gamma_P(\pi_H \to \rho \pi) = \Pi_{\cal H}$ & -\\
$\Gamma_P(\pi_{1H} \to \rho \pi) = \frac{1}{4} \Pi_{\cal H}$ & -\\
$\Gamma_P(\pi_{2H} \to \rho \pi) = \frac{1}{4} \Pi_{\cal H}$ & -\\
\end{tabular}
\begin{tabular}{l}
$\Sigma_{\cal H} \equiv g^2 \frac{|\vec{q}|^3}{m^2} \frac{b}{\pi^2 m^2} 
\left|\tfrac{2}{\pi}\langle j_1 \rangle - \langle j_0 \rangle \right|^2 $ \\ 
$\Delta_{\cal H} \equiv g^2 \frac{|\vec{q}|^3}{m^2} \frac{b}{\pi^2 m^2} 
\left|\tfrac{4}{\pi}\langle j_1 \rangle - \langle j_2 \rangle \right|^2 $ \\
$\Pi_{\cal H} \equiv g^2 \frac{|\vec{q}|^3}{m^2} \frac{b}{\pi^2 m^2} 
\left|\langle j_1 \rangle \right|^2 $ \\ 
\end{tabular}
\caption[Hybrid meson $\pi V$ decay widths]{$\pi V$ decay widths for $p=1$ light-quark hybrid mesons in the pion emission/flux-tube model. \label{pionhyb}}
\end{table}

\begin{figure} 
\begin{center} 
  \psfragscanon 
\psfrag{ma}[]{$m_{\cal H}$}
\psfrag{j0}[]{$\langle j_0 \rangle_{\mathrm{IP}}$}
\psfrag{j1}[]{$\langle j_1 \rangle_{\mathrm{IP}}$}  
\psfrag{j2}[]{$\langle j_2 \rangle_{\mathrm{IP}}$}
\psfrag{j0b}[]{$\langle j_0 \rangle_{\beta=0.4}$}
\psfrag{j1b}[]{$\langle j_1 \rangle_{\beta=0.4}$}  
\psfrag{j2b}[]{$\langle j_2 \rangle_{\beta=0.4}$}             
  \includegraphics[width=3.5in]{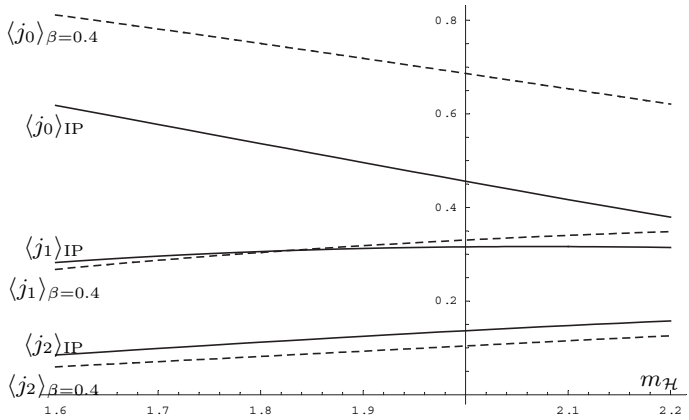}
\caption[Pion emission form-factors]{Transition matrix elements of the spherical Bessel functions, $\langle R_{\rho}(r) | j_L(|\vec{q}| r/2) | R_{\cal H} (r) \rangle$, using wavefunction parameterisations as described in the text. \label{sphbessel}}  
\end{center}
\end{figure}

We show in Figure \ref{sphbessel}, $\langle j_L \rangle= \langle {\cal C}| j_L | {\cal H}\rangle$ as a function of
$m_{\cal H}$ for ``IP'' and $\beta=0.4$ wavefunctions. Note that only
$\langle j_0 \rangle$ differs considerably between the two
wavefunction choices and as such we expect that while $P$ and $D$-wave
predictions will be rather robust with respect to wavefunction
parameterisations, $S$-wave rates will be quite  sensitive. We will
quote rates predicted with both wavefunction choices where they differ
considerably, and we will use $g=3$ throughout.

\subsection*{Negative Parity Hybrids}
\begin{itemize}

\item[$\mathbf{1^{\boldsymbol{-}\boldsymbol{-}}}$] The spin-singlet negative parity hybrid, $\rho_H \to \pi \omega$ and
may be compared with the states $\rho(1460)$ and
$\rho(1700)$\cite{Hagiwara:2002fs}, which have been suggested to have hybrid
vector meson content\cite{vector}.
 
\begin{tabular}{c|ccc} 
&  $\Gamma$ / MeV & b.r. \\ 
\hline 
$\rho(1460) \to \omega \pi$ & 29  & 9 \% \\ 
$\rho(1700) \to \omega \pi$ & 93 & 33 \% \\ 
\end{tabular} 
 
Although $\omega \pi$ decays are seen for these states, the branching
fractions have not been accurately determined. If the states are
mixtures of hybrid and conventional, as proposed in \cite{vector}, the $\pi \omega$ widths become rather model
dependent and no clear conclusions can be drawn.
 
\item[$\mathbf{1^{\boldsymbol{-}\boldsymbol{+}}}$] 
\begin{equation}\label{pi1}
  \Gamma(\pi_1(1600) \to \rho \pi) \approx 57 \mathrm{MeV}
\end{equation}
This width for the $1^{-+}$ corresponds to a branching ratio\footnote{where the error is just that from experiment.} $\sim 34 \substack{+6 \\-17}
\%$ if we identify with the state in \cite{e852} with $m=1600 \mathrm{MeV}$ and $\Gamma= 168 \pm 20 \substack{ +150 \\ -12} \mathrm{MeV}$.  This is
encouraging; had the branching ratio been $\sim 1\%$ it would be
implausible for the state to have been seen in this mode; conversely
had the branching ratio been predicted to be $\sim 100\%$ the required
absence of other channels would have disagreed with the experimental observation of this state in various channels. This result is
consistent within errors with:
  
\begin{enumerate}
\item an experimental limit $b.r.(1^{-+}(1600) \to \pi\rho) \leq
40\%$\cite{Zielinski:1987gr};
\item  with the relative branching ratios of ref.\cite{Dorofeev:1999th}:\\
$br(b_1\pi):br(\eta'\pi):br(\rho \pi) = 1:1.0 \pm 0.3:1.6 \pm 0.4$;
\item with an analysis of the E852 data, assuming purely $\rho$ exchange in the production mechanism which gave a branching ratio of $20 \pm 2\%$\cite{Page:1997xs}.
\end{enumerate}

This is an appropriate point to compare with the flux-tube breaking
model, where $\rho \pi$ decays come about only when one allows
different radial wavefunctions for the $\rho$ and the $\pi$. In
\cite{CP95} the authors consider a particular realisation of this
symmetry breaking and find a width for $\pi_1 \to \rho \pi$ of only 8 MeV,
where this assumes the hybrid is at 2 GeV; a state at 1600 MeV will
have reduced phase-space and a further reduction in width.

\item[$\mathbf{0^{\boldsymbol{-}\boldsymbol{+}}}$]
\begin{equation}
  \Gamma(\pi(1800) \to \rho \pi) \approx 480 \mathrm{MeV} \nonumber
\end{equation}
 In this model the pseudoscalar hybrid has a width significantly larger
 than the $1^{-+}$ state. We will show later that
 this is a rather general prediction of the flux-tube model. This
 numerical prediction, which would signal a very broad state indeed,
 may be better considered an upper limit. When considering conventional decays in this model we
 found that with $\beta=0.4$ wavefunctions there was little difference in overall fit quality
 between $g=2,3$. Using $g=2$ here would reduce
 the partial width to $\sim 210$ MeV.

Unfortunately we cannot really use our result to make any statement
about the hybrid character or otherwise of the $\pi(1800)$ state. As
well as our considerable theoretical uncertainties, we have no
experimental measurement of the state's $\rho \pi$ branching
ratio. The experiment that we would look to for such data is E852 but
they note in \cite{Chung:1999we} that the pseudoscalar partial wave suffers
badly from the required $(\pi\pi)_S$ parameterisation uncertainties
and as such no reliable data can be extracted.

\item[$\mathbf{2^{\boldsymbol{-}\boldsymbol{+}}}$] The isovector $2^{-+}$ state in this model has the same partial width
as the $1^{-+}$ state and hence should be as prominent in
experiment. There is a very broad ($\Gamma \sim 600$ MeV) candidate
$\pi_2$ state seen at 2100 MeV in $\rho \pi$ and $f_2 \pi$
modes\cite{Hagiwara:2002fs}, which may correspond to the broad,
unparameterised enhancement in $\rho \pi$ above 2 GeV seen by E852
\cite{Chung:1999we}. The observation in $\omega \rho$ of
$\pi_2(1950)$\cite{Lu:2004yn} casts some doubt over a hybrid
interpretation.

\end{itemize}
\subsection*{Positive Parity Hybrids}

\begin{itemize}

\item[$\mathbf{1^{\boldsymbol{+}\boldsymbol{+}}}$]
\begin{align}\label{axial}
  \Gamma_S(a_{1H}(2100) \to \rho \pi) &\approx \begin{matrix} 160 \\
    660\end{matrix} \;\mathrm{MeV}\nonumber \\
  \Gamma_D(a_{1H}(2100) \to \rho \pi) &\approx \begin{matrix} 110 \\
    170\end{matrix} \;\mathrm{MeV} \nonumber
\end{align}
The upper/lower values are with ``IP'', $\beta=0.4$ wavefunctions and
with $g=3$. We can probably consider these values to be lower and
upper limits on the $S$-wave width - the $\beta=0.4$ wavefunctions are
the optimum choice for $L=1 \to L=0$ conventional transitions, the
``IP'' solution shows that hybrids have a slightly smaller $\beta$
than $L=1$ states and hence we expect $\langle j_0 \rangle$ to fall
faster with $|\vec{q}|$ than in the $\beta =0.4$ case. The effective
$\beta$ in the ``IP'' solution is too small in the conventional sector
and hence probably too small for hybrids too - hence our upper/lower
assignment for $\beta=0.4$, ``IP''. Furthermore it has been noted in \cite{Close:2003tv} that the quark-model has a tendency not to
describe well decays in which the end state is in an $S$-wave, while
the quarks in the initial meson are in a higher angular momentum
eigenstate.

What is clear is that we have a considerable partial width into $\rho
\pi$ for the axial hybrid. This is at odds with the claim that the
$a_1(2096)$ state seen by E852 has hybrid character as it is not seen
at all in $\rho \pi$ by E852, who see only the dominant $a_1(1260)$ in
$1^{++}$.

\item[$\mathbf{2^{\boldsymbol{+}\boldsymbol{-}}}$] In the positive parity sector there are exotics $(0,2)^{+-}$. The
spin-0 state has no decay into $\pi V$, but we have some hope of
seeing the spin-2 state if this model is correct. Normalising against
the exotic $1^{-+}$ we have,
\begin{equation}\label{}
  \frac{\Gamma(b_{2H} \to \omega \pi)}{\Gamma(\pi_{1H} \to \rho \pi)}
  \approx \frac{72}{5 \pi^2}
  \left(\frac{|\vec{q}_2|}{|\vec{q}_1|}\right)^3 \left( 1-
    \frac{\pi}{4}\frac{\langle j_2 \rangle}{\langle j_1
      \rangle}\right)^2,\nonumber
\end{equation}
where we've neglected the slow mass dependence of $\langle j_{1,2}
\rangle$. This ratio suggests a similar partial width for the
two states.

Normalising against the conventional $2^{++}$ decay gives
\begin{equation}\label{}
  \hspace{1cm}\frac{\Gamma(b_{2H} \to \omega \pi)}{\Gamma(a_2 \to \rho \pi)} = \left(\frac{|\vec{q}_{\cal H}|}{|\vec{q}_{\cal C}|}\right)^3 \frac{4 b}{\pi^3 m^2} \left| \frac{\langle j_1 \rangle_{\cal H} - \tfrac{\pi}{4} \langle j_2 \rangle_{\cal H}}{\langle j_1 \rangle_{\cal C} - \frac{\langle\overleftarrow{\partial}  j_2  \rangle_{\cal C} }{ m } } \right|^2\nonumber
\end{equation}
where there is some suppression for hybrids from $\frac{4 b}{\pi^3
  m^2} \approx 0.2$ but which is compensated by the increase in phase
space. For a $b_{2H}$ at 1600 MeV we anticipate a partial width around
$50 \%$ of the $a_2 \to \rho \pi$ partial width. A heavier $b_{2H}$ at
2100 MeV, with the increase in phase space would be around $150 \%$ of
$a_2$. Thus we expect $b_{2H} \to \omega \pi$ with a partial width of
tens of MeV.

The possible sighting of such a state around 1650 MeV by E852 in
$\pi^- p \to (\omega \pi^-) p$\cite{Nozar:2002br} is, in light of our
estimates, rather interesting and a dedicated study of this
observation would be enlightening.
\end{itemize}

\section*{Higher order effects}  \label{sect:caveat}      

The reader will recall that in equation \eqref{hybridcurrent} we chose
to neglect a term transforming as $\vec{\sigma}.\vec{p}' \;
\vec{q}.\vec{a}$ on the grounds that it is sub-leading by one power of
$|\vec{q}|/m$, or equivalently $v/c$. Unfortunately, for the states we
are considering, $|\vec{q}|/m$ is not necessarily small and our
truncation appears artificial. A common approach in quark model
treatments of hadron decays is to truncate the non-relativistic
expansion of the operator at the highest order in $v/c$ for which we
know all possible terms. We do not include the subset of effects at
the next order that we are able to calculate and they may be negated
by other effects at this order that we are unable to calculate.

Explicit calculation of the ``suppressed'' $\vec{\sigma}.\vec{p}' \;
\vec{q}.\vec{a}$ term shows that it has a considerable effect on our
predictions, especially in the negative parity sector. Its net effect
is to modify the $P$ and $D$-wave amplitudes according to
\begin{align}
\Pi_{\cal H} \to \Pi_{\cal H}' &=  g^2 \frac{|\vec{q}|^3}{m^2} \frac{b}{\pi^2 m^2} \left| \langle j_1 \rangle + \frac{2}{\pi} \frac{\langle \overleftarrow{\partial} j_1  \rangle}{m} \right|^2 \nonumber  \\
\Delta_{\cal H} \to  \Delta_{\cal H}' &= g^2 \frac{|\vec{q}|^3}{m^2} \frac{b}{\pi^2 m^2}  \left| \frac{4}{\pi}\langle j_1 \rangle - \langle j_2 \rangle - \frac{6}{\pi}   \frac{\langle \overleftarrow{\partial} j_2  \rangle}{m} \right|^2.\nonumber
\end{align}
With harmonic
oscillator wavefunctions $\overleftarrow{\partial} \to - \beta^2_{\cal
  C} r$ and $\Pi_{\cal H}' \sim \left| \left\langle \left(1- \frac{2 \beta^2_{\cal C} r}{\pi m}\right)j_1 \right\rangle\right|^2$ which is approximately zero for light quarks due to an accidental cancellation. In \cite{thesis}, non-adiabatic corrections to this simplest flux-tube model are investigated and this approximate zero does not survive, hence we cannot trust that the cancellation is physically
relevant. Unfortunately this also means that we must associate a
considerable theoretical error with our predictions.

In light of this disturbing sensitivity to the order of truncation
used, we should ask if there are any {\em more general} results to be
extracted from this study. We find that there are and we discuss them
in the next section.
 
\section*{General current structure arguments}\label{sect:structure}

Making only the assumptions that the pion current should
transform as a pseudoscalar, be linear in $\vec{a}$ and at most first
order in $\vec{\sigma}$ we can have only the following possible structures,
\begin{align}\label{}
  j_\pi({\cal P_H}=+) &= \alpha \sigma_z \vec{a}.\hat{z} + \beta [
  \sigma_- \vec{a}.\hat{x}_+ + \sigma_+ \vec{a}.\hat{x}_-]\nonumber\\
 j_\pi({\cal P_H} = - ) &= \gamma[
  \sigma_- \vec{a}.\hat{x}_+ + \sigma_+ \vec{a}.\hat{x}_-].\nonumber
\end{align}
This form is consistent with that following from the Melosh transformation for $\pi$ induced transitions\cite{meloshpi}. As such it is more general, though less predictive, than any particular model where specific values for the parameters $\alpha,\beta,\gamma$ also would obtain\cite{adamsz}.
Coupling the internal $L,S$ by the Clebsch-Gordan coefficient,
$\langle L\!\!\!=\!\!\!1, m'; S, m_S | J,m_J \rangle$ and using the flux-tube matrix elements from \cite{CD03b},  we find helicity amplitudes\footnote{Unless $\rho$ or $\omega$ is specifically denoted,
  our amplitudes for $A \to \pi V$ refer to a single charge mode for
  $\pi$ and $V$. Widths for $I=(0;1) \to \pi (\rho$ or $\omega)$ then
  require relevant charge counting factors to be included}

\begin{center} 
\begin{tabular}{c|cc || c|cc} 
${\cal P}=-$ & ${\cal M}_0$ & ${\cal M}_\pm$ &$ {\cal P}=+$ & ${\cal M}_0$ & ${\cal M}_\pm$ \\ 
\hline 
$1^{--}$ & 0 & $\mp \sqrt{2} \gamma$ &$1^{++}$ & $\alpha$ & $- \sqrt{2} \beta$ \\ 
$0^{-+}$ & $\sqrt{\frac{8}{3}} \gamma$ & 0& $0^{+-}$ & 0 & 0 \\ 
$1^{-+}$ & 0 & $\pm \gamma$ & $1^{+-}$ & $2 \beta$ & $- \frac{1}{\sqrt{2}} \alpha + \beta$ \\ 
$2^{-+}$ & $\frac{2}{\sqrt{3}} \gamma$ & $\gamma$ & $2^{+-}$ & 0 & $\pm (\frac{1}{\sqrt{2}} \alpha + \beta)$ 
\end{tabular} 
\end{center}

Using the conversion from helicity to partial-wave amplitudes in Table
XI of \cite{Godfrey:1985xj},  these can be succinctly expressed as follows,

\begin{center} 
\begin{tabular}{c|cccc |c} 
${\cal P}=+$& $1^{++}$ & $0^{+-}$ & $1^{+-}$& $2^{+-}$ &\\ 
\hline 
$A_S=$ & $\sqrt{3}$ & 0 & $\sqrt{6}$ & 0 & $\times S$ \\  
$A_D=$& $\sqrt{6}$ & 0 & $\sqrt{3}$ & 3 & $\times D$  
\end{tabular} 

\vspace{.5cm}

\begin{tabular}{c|cccc |c} 
${\cal P}=-$& $1^{--}$ & $0^{-+}$ & $1^{-+}$& $2^{-+}$ & \\ 
\hline 
$A_P=$ & $\frac{3}{\sqrt 2}$ & $\sqrt 3$   & $\frac{3}{2}$ & 
$\frac{\sqrt{15}}{2}$ & $ \times P$ 
\end{tabular} 
\end{center}
where $S=\frac{1}{3}(\alpha - 2\sqrt{2} \beta)$,
$P=\frac{\sqrt{8}}{3}\gamma$ and $D=-\frac{1}{3}(\alpha + \sqrt{2}
\beta)$.
 
These correlate with the relative amplitudes in Table 6 of \cite{CP95} who computed these decays in the assumption that the
flux tube breaks with creation of a new $q\bar{q}$ in $^3P_0$ state. 
 
With the assumption only that the partial wave amplitudes $S,D; (P)$
are common to the supermultiplets of hybrids with ${\cal P} = +(-)$
respectively, then for equal masses where $\Gamma \sim \frac{C}{2J+1}
\sum_L { |A_L|}^2$ we have the following constraint on the widths for
the isoscalar/isovector ${\cal P} = +$ states to $\pi V$.  
\begin{multline}
3\Gamma \left(1^{+-}\left\{ I=0
      \to \rho \pi \right\}\right) + 5\Gamma
\left(2^{+-}\left\{I=0 \to \rho
      \pi  \right\}\right) \\
= 9\Gamma
\left(1^{++}\left\{ I=1 \to \rho
      \pi \right\}\right) \nonumber
\end{multline} 
For the ${\cal P}=-$ states
we have, 
\begin{multline}
 \left(\Gamma(1^{-+}) =\Gamma(2^{-+})
  =\frac{1}{4}\Gamma(0^{-+})\right)\left\{I=1 \to \rho \pi  \right\}
\\
=\frac{1}{3}\Gamma(1^{--})\left\{I=0 \to \rho \pi  \right\}. \nonumber
\end{multline} 
It is trivial to check that the expressions we have derived explicitly
satisfy these rules. As we have already mentioned, the predictions of
Close and Page\cite{CP95} satisfy these rules, as do the predictions of
an alternative flux-tube breaking model\cite{Page:1998gz} (which has different
quantum numbers at the breaking point). These ``sum-rules'' appear to
be a rather general property depending only upon the spin-orbit
coupling structure of the states, consequently we expect them to be good if
the flux-tube model is a good description of hybrid meson structure.

Their practical use is that once we have a candidate in $(\rho/\omega) \pi$ we
can estimate the partial widths of other hybrid states - even the non-exotic
ones - in a relatively model-independent way. For example if the
$\pi_1(1600)$ is a hybrid, then $\pi_{0H} \to \pi\rho$ must also be
prominent.

\section*{Radiative decays by Vector Meson Dominance}

For the conventional hadrons, the widths into $\pi V$ may be used to
give estimates for the widths into $\pi \gamma$ by converting the $V
\to \gamma$ as in vector dominance. The basic premise is that the
photon has some hadronic character - it can fluctuate into an
off-shell vector meson with some amplitude and interact strongly with
the hadron target. Our phenomenological treatment will follow that of
Babcock \& Rosner\cite{Babcock:1976hr}. We find satisfactory agreement
with experiment in the conventional sector for the radiative decays of the
$a_{1,2}, b_1$\cite{thesis}, which motivates our application of the
method to hybrid states.
 
For the exotic hybrid candidate $\pi_1(1600)$, using the $\rho \pi$ partial
width prediction of 57 MeV we find a $\gamma \pi$ partial width of
$\sim 170$ keV. This is a healthy width, comparable to the
conventional $b_1 \to \gamma \pi$ width of $230 \pm 60$ keV. The
expectation of the flux-tube breaking model supplemented with VMD is
of a maximum width of $\sim 70$ keV with a more realistic prediction
of $20 \%$ of this\cite{Close:1995pr}.

The non-exotic $2^{-+}$ hybrid, according to the ``sum rules'' of the previous section will have a partial width equal to this
with modification only for the potentially different state mass.

The exotic $b_2$, if it has a mass $\sim 2100$ MeV has $\gamma \pi$
width $\sim 50$ keV. Much of the suppression relative to the $\pi_1$
width is down to the factor of $1/9$ caused by $g_\omega = 3
g_\rho$. This does not appear for the isosinglet $2^{+-}$ hybrid and
there we expect $\Gamma(h_2(2100) \to \gamma \pi) \sim 450
\mathrm{keV}$. 

The axial hybrid $a_{1H}$ was predicted to have a potentially very large $\rho \pi$
width which was rather sensitive to wavefunction
parameterisation. This large width unsurprisingly leads in VMD to a
large radiative width $\sim 550 \to 1600$ keV which is comparable with
our prediction in \cite{CD03a, CD03b} on the basis of an $E1$ photon
current exciting the flux-tube in a pion. 

\subsubsection*{Photoproduction of hybrids through pion exchange}

At high photon energies and low momentum transfer, $t$,
photoproduction of mesons is believed to have a significant contribution from the pion
exchange mechanism. There is very little data available, but what
exists is consistent with one pion exchange expectations (see
e.g. \cite{Afanasev:1999rb}). If this is really a significant mechanism
then we should expect the hybrids we have identified as having
large $\gamma \pi$ partial widths to be produced prominently in
photoproduction.

In particular we noted that an isoscalar $2^{+-}$ would be large in
$\gamma \pi$. A recent analysis of photoproduction of
$\pi^+\pi^-\pi^0$ by the CLAS collaboration\cite{adamstalk} finds a large
bump in the $2^{+-}$ partial wave near 2 GeV. Given that we also
expect this state to have a partial width into $\pi \rho$ comparable
to $\Gamma(a_2 \to \rho \pi)$ this signal is most interesting. 

As well as photoproduction of isoscalar/isovector $2^{+-}$ in pion exchange, there is also
the possibility of diffractive photoproduction where the photon
fluctuates into a $\omega/\rho$ which fuses in a $P$-wave with the
Pomeron. We don't currently have the tools to calculate a rate for
this process, but we have no reason to expect it to be small - the
vector meson is already in the required spin-triplet so the Pomeron current
could either excite the flux-tube by oscillating the quark or by
interacting directly with the tube.

Hence the $2^{+-}$ exotic should be especially favoured in
photoproduction: the $I=0$ coming from $\pi$ exchange and the $I=1$
from Pomeron/gluon exchange.

For the case of pion exchange dominance, a study similar to that in \cite{Afanasev:1999rb}, using the full
theoretical formalism of photoproduction modeling could be carried out
using the matrix elements found in this model.

\section*{Summary}

We have used the formalism in which currents acting on quarks can
excite the flux-tube to consider the pionic decays of hybrid
mesons. The pion is considered to act as a pointlike current which
couples only to the quarks (isospin ensures that it cannot couple
directly to the flux-tube) and reasonable success in describing
conventional meson decays is observed.

Previous studies of hadronic hybrid meson decays had assumed the need
for pair production at some point on the flux-tube and had found that
decay to a pair of mesons with identical spatial wavefunctions were
forbidden. In the quark model with no spin-dependent effects, the pion
and the rho have identical wavefunctions and this has lead to the
expectation that the $\pi \rho$ channel for hybrid decay should be
suppressed. The model outlined here maximally breaks the
$\pi/\rho$ symmetry and finds that $\pi\rho$ rates can be large, which
seems to be required by the data showing a $1^{-+}$ resonance in the
$\pi\rho$ channel.

Detailed consideration of the model shows that the $v/c$ expansion of
the pionic current is not under control and hence that numerical
predictions may not be robust. In light of this, more general results
were extracted which link $\pi\rho$ partial widths of different hybrid
states and which appear to be dependent only on the spin-orbit
structure of the hybrid meson states.

Radiative decay widths of hybrids were discussed under the assumption
of VMD converting $\rho \to \gamma$ and were found to be not
necessarily small. This offers hope to the experiments intending to
produce hybrids via photoproduction, especially for $J^{PC}=2^{+-}$.

\begin{center} {\bf Acknowledgments} \end{center}
  
This work is supported, in part, by grants from the Particle Physics
and Astronomy Research Council, and the EU-TMR program ``Euridice'',
HPRN-CT-2002-00311.

\end{document}